# Valley Magnetoelectricity in Single-Layer MoS$_2$

Jieun Lee[1,2], Zefang Wang[1], Hongchao Xie[1], Kin Fai Mak[1]*, Jie Shan[1]*

[1]Department of Physics and Center for 2-Dimensional and Layered Materials,
The Pennsylvania State University, University Park, Pennsylvania 16802-6300, USA
[2]Department of Physics and Energy Systems Research, Ajou University, Suwon, Gyeonggi-do 16499, Korea

Correspondence to: kzm11@psu.edu (K.F.M.); jus59@psu.edu (J.S.)

**Magnetoelectric (ME) effect, the phenomenon of inducing magnetization by application of an electric field or vice versa, holds great promise for magnetic sensing and switching applications [1]. Studies of the ME effect have so far focused on the control of the electron spin degree of freedom (DOF) in materials such as multiferroics [2] and conventional semiconductors [3]. Here, we report a new form of the ME effect based on the valley DOF in two-dimensional (2D) Dirac materials [4-6]. By breaking the three-fold rotational symmetry in single-layer MoS$_2$ via a uniaxial stress, we have demonstrated the pure electrical generation of valley magnetization in this material, and its direct imaging by Kerr rotation microscopy. The observed out-of-plane magnetization is independent of in-plane magnetic field, linearly proportional to the in-plane current density, and optimized when the current is orthogonal to the strain-induced piezoelectric field. These results are fully consistent with a theoretical model of valley magnetoelectricity driven by Berry curvature effects. Furthermore, the effect persists at room temperature, opening possibilities for practical valleytronic devices.**

Electrons in two-dimensional (2D) Dirac materials including gapped graphene and single-layer transition metal dichalcogenides (TMDs) possess a new two-fold valley degree of freedom (DOF) corresponding to the K and K' valleys of the Brillouin zone. The valley DOF carries orbital magnetic moment [4-6]. A net valley magnetization forms the basis for valley-based applications. Such magnetization can arise from either a finite population imbalance between the valleys (i.e. a net valley polarization) or a distribution difference between them without a population imbalance [5]. Whereas the former relaxes by intervalley scattering, the latter is largely limited by intravalley scattering. The presence of valley contrasting Berry curvatures in 2D Dirac materials, which can couple to external electromagnetic excitations, enables the control of valley magnetization [6-17]. Although the control of valley magnetization by circularly polarized light and by a vertical magnetic field has now become routine [6-8, 10-16], the development of practical valleytronic devices requires the pure electrical control of valley magnetization. The valley magnetoelectric (ME) effect is an attractive approach for this purpose.

To realize the linear ME effect, a material has to possess broken time-reversal and spatial-inversion symmetries [1, 18]. For an electrical conductor, time-reversal symmetry can be broken naturally by application of a bias voltage, under which dissipation through carrier scattering is caused by a charge current. The magnetoelectricity produced in this manner is known as the kinematic ME effect [19, 20]. Single-layer TMDs such as MoS$_2$, which are intrinsically non-centrosymmetric, satisfy these symmetry requirements with finite doping and biasing. The characteristic three-fold rotational symmetry of these materials, however, renders the relevant valley ME susceptibilities zero [18, 21]. In this work, we apply a uniaxial stress to lower the crystal symmetry and demonstrate the generation of steady-state out-of-plane valley magnetization by an in-plane charge current in single-layer MoS$_2$. Such an effect can be understood intuitively as the valley analog of the Rashba-Edelstein effect for spins [20, 22, 23] (Fig. 1b). In the presence of an in-plane electric field $\mathcal{E}_{pz}$, an in-plane current density $J$ generates an effective magnetic field $\propto \mathcal{E}_{pz} \times J$, resulting in finite out-of-plane valley magnetization $M_V$. In place of the Rashba field (an out-of-plane built-in electric field at the semiconductor heterostructure interfaces) in the Rashba-Edelstein effect [20, 22, 23], $\mathcal{E}_{pz}$ here is the in-plane piezoelectric field that arises from the strain-induced lattice distortion and the associated ion charge polarization (Fig. 1a), a phenomenon that has been recently demonstrated in single-layer MoS$_2$ [24, 25]. The effect provides a possible valley magnetization-current conversion mechanism for magnetization switching and detection applications [22, 23].

In our experiment, single-layer MoS$_2$ was mechanically exfoliated from bulk crystals onto flexible polydimethylsiloxane (PDMS) substrates. The flakes were stressed along a particular high-symmetry axis by mechanically stretching the substrate. The strained flakes were then transferred onto SiO$_2$/Si substrates with pre-patterned electrodes to form field-effect transistors with Si as the back gate. Strain was fixed for all subsequent measurements. During the process, the strain level in the MoS$_2$ flakes was monitored by the photoluminescence (PL) energy shift based on the strain-induced decrease of the band gap (~ 20 nm/% strain independent of direction [26, 27]) (Fig. 1e). The crystallographic orientation of the flakes was determined by the second-harmonic generation (SHG) under normal incidence [26]. Figure 1d shows the intensity of the second-harmonic component parallel to the excitation as a function of the polarization angle, the maximum of which corresponds to the armchair direction. Polar magneto-optic Kerr rotation (KR) microscopy at variable temperatures was employed to image the spatial distribution of the magnetization in the device channels [17]. To this end, linearly polarized light was focused onto the device under normal incidence and its polarization rotation (i.e. KR angle) upon reflection was detected. The polar configuration is sensitive to out-of-plane magnetization only. To enhance the detection sensitivity, we have used a probe wavelength close to the MoS$_2$ A exciton resonance and a lock-in detection method with the bias voltage being modulated at 4.11 kHz. The KR sensitivity is about 0.5 μrad/Hz$^{1/2}$ (details see Methods).

In Figs. 2a and 2b, we compare the KR image of two single-layer MoS$_2$ devices at 30 K, one with intentional strain and one without. Drastic differences are observed. In the unstrained device (Fig. 2a), the KR is present only near the channel edges and is of opposite sign on the two edges. When the current direction is reversed, the KR switches sign on the two edges. (Detailed studies on the gate and bias dependences are shown in Supplementary Sect. 1.) Such a KR signal is originated from valley polarization accumulated on the channel edges that is driven by the valley Hall effect (VHE), as reported by recent experiments [8, 9, 17]. On the other hand, in the strained device (Fig. 2b), a much stronger KR signal per unit current (supplementary Fig. S2) is observed nearly uniformly over the entire device channel. The signal also changes sign under a reversed current direction. For this particular device, the sample was strained by ~ 0.2% along the zigzag direction and biased along the same direction.

To further examine the observed KR signal in the bulk of strained single-layer MoS$_2$, we performed multiple checks (details see Supplementary Sect. 4). First, KR was measured as a function of the probe beam's polarization direction. The observed negligible dependence excludes strain-induced optical birefringence as a possible origin of the observed KR. Second, we studied the dependence of the KR signal on the channel current density ($J$). Figure 2c shows the KR (symbols) and the current density (solid lines) as a function of gate voltage ($V_g$), as well as bias voltage ($V_{ds}$) (inset of Fig. 2c). A typical *n*-type transport behavior is seen. The observed KR is linearly proportional to $J$. Third, we studied the effect in the presence of an in-plane magnetic field in the Voigt geometry. No dependence of the KR signal on the magnetic field is observed up to $\pm$ 0.5 T (Fig. 2d). The absence of the Hanle effect (i.e. Larmor precession of spin magnetization under an in-plane magnetic field [28]) excludes the spin nature of the observed effect in case of weak spin-orbit coupling (SOC). It is, however, fully compatible with valley magnetization, which does not couple with an in-plane magnetic field [29].

To understand the microscopic mechanism of the observation, we consider the effect of strain on the Dirac Hamiltonian of single-layer MoS$_2$ in the k·p approximation [4, 30, 31], $H_0 = \hbar v_F(k_x\sigma_x + sk_y\sigma_y) + \frac{\Delta}{2}\sigma_z$. Here $\hbar$ and $v_F$ denote, respectively, the Planck's constant and the Fermi velocity, $\hbar\boldsymbol{k} = \hbar(k_x\hat{x} + k_y\hat{y})$ is the crystal momentum measured from the K ($s$ = +1) and K' ($s$ = -1) point of the Brillouin zone, $\boldsymbol{\sigma}$ is the Pauli matrix for the sublattice index, and $\Delta$ is the band gap. We choose unit vector $\hat{x}$ and $\hat{y}$ along the armchair and zigzag direction, respectively ($\hat{z}$ corresponds to the out-of-plane direction). The SOC has been ignored since we only consider *n*-type MoS$_2$ and the SOC is weak in the lowest-energy conduction bands of MoS$_2$ [32]. Under uniform uniaxial strain along high-symmetry axes, the Hamiltonian is modified as $H \approx \hbar(v_{Fx}k'_x\sigma_x + sv_{Fy}k'_y\sigma_y) + \frac{\Delta'}{2}\sigma_z + s\alpha(u_{xx} - u_{yy})\hbar v_F k'_y$ [30, 31]. Here the crystal momentum $\hbar\boldsymbol{k}$ is replaced by the canonical momentum $\hbar\boldsymbol{k}'$ due to a strain-induced fictitious vector potential, and a new (third) term that does not couple to the Pauli matrix appears with $\alpha \sim 1$ and $u_{ij}$ denoting, respectively, a dimensionless parameter and a strain tensor element [30, 31]. The Hamiltonian also takes into account the anisotropy in the Fermi velocity due to broken three-fold rotational symmetry ($v_{Fx} \neq v_{Fy}$) and the shift of the energy gap from $\Delta$ to $\Delta'$. The major effect of these changes is the shift of the band extrema and the extrema of the Berry curvature distribution (blue dashed lines) illustrated schematically in Fig. 1c. The band extrema are shifted from the original K/K' point (black dashed lines) to opposite directions, which also do not coincide with the extrema of the Berry curvature distribution. Under a bias field, a current is formed and the Fermi surface tilts to the direction of the bias field. Although the two valleys maintain an equal population, the modified band structure and Berry curvature distribution, in combination with the non-equilibrium carrier distribution, lead to a net valley magnetization:

$$\boldsymbol{M_V} \approx \frac{3\alpha\varepsilon\hbar v_F}{e_{xx}\Delta'}\boldsymbol{\mathcal{E}_{pz}} \times \boldsymbol{J}. \qquad (1)$$

Here $\varepsilon$ and $e_{xx}$ denote, respectively, the dielectric constant and the piezoelectric tensor element of single-layer MoS$_2$, and only the leading-order term in both the piezoelectric field $\boldsymbol{\mathcal{E}_{pz}}$ and the current density $\boldsymbol{J}$ is considered. Similarly, the maximum absorbance difference for left- and right-handed light (which determines the KR angle) is calculated by ignoring the excitonic effect as $A_+ - A_- \approx M_V A \frac{\Phi_0}{3\Gamma}$ with $\Phi_0$, $A$ and $\Gamma$ denoting, respectively, the magnetic flux quantum, the peak absorbance and the width of the optical transition (See Supplementary Sections 5 and 6 for detailed derivations).

Equation (1) reproduces our intuitive result for the valley ME effect in single-layer TMDs as illustrated in Fig. 1b. It is also consistent with the symmetry requirements [18, 21]. We consider the ME susceptibility $\alpha_{ij}$ that relates the bias electric field $\mathcal{E}_j$ to magnetization $M_i = \alpha_{ij}\mathcal{E}_j$. Since single-layer MoS$_2$ does not spontaneously break time-reversal symmetry, a charge current under bias that causes dissipation is required for finite $\alpha_{ij}$ [19, 20]. For unstrained single-layer MoS$_2$ with $D_{3h}$ symmetry, the ME susceptibilities $\alpha_{zx}$ and $\alpha_{zy}$ that are relevant for the generation of out-of-plane magnetization by an in-plane bias field all vanish [18, 21]. However, under a uniaxial stress along the armchair or the zigzag direction, the symmetry point group is reduced to $C_{2v}$ with the mirror line along the $\hat{x}$ (armchair) axis. The $\alpha_{zy}$ component becomes nonzero, and finite magnetization $M_z = \alpha_{zy}\mathcal{E}_y$ emerges for a bias field with nonzero component perpendicular to the mirror line (i.e. $\mathcal{E}_y \neq 0$) [18, 21].

Equation (1) also explains our major experimental findings shown in Fig. 2, namely, the need of finite strain to generate valley magnetization and the linear dependence of the KR on the amplitude and sign of the current density. To further verify the vector relation of $\boldsymbol{M_V} \propto \boldsymbol{\mathcal{E}_{pz}} \times \boldsymbol{J}$, we fabricated devices in the Corbino disk geometry, which allows current along the radial direction over a wide range of angles relative to the applied stress. Two devices are shown in Fig. 3 with a uniaxial stress applied along the armchair (upper row) and zigzag axes (lower row). The piezoelectric field $\boldsymbol{\mathcal{E}_{pz}}$ is along the $\hat{x}$-axis with opposite direction in the two cases (solid orange lines). Note the effect of tensile strain along the zigzag direction is similar to compressive strain in the armchair direction. Both devices show positive KR in one half and negative KR in the second half of the channel (Figs. 3b and 3e). In particular, the KR signal vanishes along the orange line and reaches maximum approximately perpendicular to it. The observed patterns agree reasonably well with the simulation result based on $M_V \propto \hat{z} \cdot (\boldsymbol{\mathcal{E}_{pz}} \times \boldsymbol{J})$ for both devices (Figs. 3c and 3f). (Details of simulation are provided in Supplementary Sect. 2.) Deviations of the experimental results from the predictions may originate from the presence of inhomogeneous strain in the samples. Inhomogeneous doping and optical responses associated with it, as well as strain-induced conductivity anisotropy may also play a role. However, the observed images with one-fold symmetry cannot be explained by either the strain-induced conductivity anisotropy or the trigonal warping effect of the band structure [7]. (See Supplementary Sections 4.4 and 4.5 for more details.) Moreover, Eqn. (1) predicts a KR signal level that is comparable to the value observed in experiment (see Methods). We thus conclude that out-of-plane valley magnetization has been generated and observed by KR in strained single-layer MoS$_2$ through the valley ME effect.

Finally, we examine the temperature dependence of the effect. The gate dependence of the two-point square conductance $\sigma$ (Fig. 4a) and the KR (Fig. 4b) from the same device is shown side by side at several selected temperatures. A clear metal-insulator crossover is observed from the transport data around $V_g \approx 12$ V. The device channel shows a metallic behavior above this gate voltage (i.e. $\sigma$ decreases with increasing temperature), and an insulating behavior below it (i.e. $\sigma$ increases with temperature). The observed square conductance at the crossover ($\sigma \approx 20$ μS) is smaller than the Ioffe-Regel criterion ($\sigma \approx 40$ μS) [33], which is likely due to an underestimated experimental value by the two-point method with Schottky contacts. The KR of the device under fixed strain shows a similar temperature dependence, as expected from $M_V \propto \mathcal{E}_{pz} \times J$. The detailed temperature dependences of the two quantities, however, are not fully identical since the parameters involved in these measurements (such as the optical conductivity at the probe wavelength, the optical resonance width and the Schottky barrier height) are also temperature dependent. Detailed studies of the temperature dependences are beyond the scope of this work. However, we draw attention to an important finding here: valley ME effect persists at room temperature. The gate dependence of the valley magnetization at room temperature is shown in the inset of Fig. 4b, and the spatial image and bias dependence, in Supplementary Sect. 3. The observation of electric-field induced valley magnetization at room temperature opens up the possibility of valley-based applications, such as magnetic switching [22, 23] and flexible devices [24, 25], by engineering the Berry curvature effects.

**Methods**

**Device fabrication**

Atomically thin flakes of $MoS_2$ were mechanically exfoliated from synthetic bulk crystals (2D Semiconductors) onto flexible polydimethylsiloxane (PDMS) substrates. Single-layer flakes were identified by the combination of their optical contrast and photoluminescence (PL) spectrum. The crystallographic orientation of each single-layer flake was determined by performing optical second-harmonic generation (SHG) on multiple locations of the flake to ensure that it contains a single domain. The flakes were then stressed uniaxially along a high-symmetry axis by stretching the PDMS substrates using a mechanical translation stage. The PL spectrum was simultaneously measured to monitor the strain level. Strain up to about 0.2 - 0.3% (corresponding to a shift in the PL peak wavelength by 4 - 6 nm) was typically achieved in monolayer $MoS_2$ before the flake started to slip or break. A glass slide was then employed to support the stretched PDMS for sample transfer onto a Si substrate with a 100-nm $SiO_2$ layer. During the transfer, the plane of the $MoS_2$ flake and the Si substrate were kept parallel so that strain was best maintained. The Si substrates were pre-patterned with electrodes (Ti 3nm/Au 30nm) by either e-beam or photo-lithography followed by metal evaporation so that the $MoS_2$ devices are free of resist contamination.

**Second-harmonic generation (SHG)**

To determine the crystallographic orientation of single-layer $MoS_2$ flakes, we employed SHG measurements based on a femtosecond Ti:Sapphire oscillator centered at 820 nm with a repetition rate of ~ 80 MHz and a pulse duration of ~ 100 fs. The laser beam was focused onto the sample under normal incidence using a 40x microscope objective. The reflected second-harmonic (SH) radiation peaked at 410 nm was filtered and detected by a spectrometer equipped with a nitrogen-cooled charge-coupled device (CCD). The excitation power was kept below 1 mW on the sample. The crystallographic orientation of the crystal was determined from the SH intensity as a function of the excitation polarization with respect to the sample's high-symmetry axes. We chose to fix the sample while varying the linear polarization of the excitation beam using a broadband half-wave plate. The reflected SH beam was passed through the same broadband half-wave plate and an analyzer for the measurement of the SH component parallel to the excitation polarization. The integration time for each SH spectrum was 5 seconds with binning applied to the CCD readout. In Fig. 2d, we show the integrated SH intensity as a function of excitation polarization. A six-fold pattern was observed as expected from $D_{3h}$ symmetry of the crystal. The SH maxima correspond to the armchair direction.

**Photoluminescence (PL) spectroscopy**
PL spectroscopy was performed on $MoS_2$ flakes to first identify their monolayer thickness and then to characterize the strain level. A continuous-wave solid-state laser at 532 nm (up to 50 µW on the sample) was employed as the excitation source. PL spectrum was collected with a confocal microscope setup, filtered, and recorded by a spectrometer equipped with a CCD.

**Magneto-optic Kerr rotation (KR) microscopy**
KR microscopy was performed on devices in an optical cryostat. A linearly polarized probe laser was tuned to a wavelength slightly red shifted from the PL peak wavelength and focused perpendicularly onto the devices by a 40x microscope objective. The power on the sample was 200 - 350 µW. The reflected probe was collected by the same objective, passed through a half-wave plate, split by a Wollaston prism, and detected by a pair of balanced photodetectors. In order to increase the signal-to-noise ratio, an oscillating bias voltage at 4.11 kHz was applied to the devices and the KR was detected using a lock-in amplifier with a time constant of 30 msec. The noise level was about 0.5 µrad/Hz$^{1/2}$ away from the boundaries of the gold electrodes. The noise level is typically higher near the boundaries due to distortions of the light polarization upon reflection. For two-dimensional (2D) spatial mapping, the devices were scanned by an XY piezo-stage while the probe beam was fixed. The KR measurement in the presence of an in-plane magnetic field was performed in a cryostat equipped with a superconducting magnet in the Voigt geometry.

The detailed analysis that relates the KR angle $\theta$ to the optical properties of the sample has been discussed in Supplementary Section 6 of Ref. [17]. In short, the KR angle is proportional to the difference in absorbance of left- and right-handed light at the probe energy: $\theta = Im[\beta_{loc}(A_+ - A_-)]$. Here $Im[…]$ denotes the imaginary part and $\beta_{loc}$ describes the combined properties of the substrate, oxide layer and the local field factors that influence wave propagation in the multilayer structure and is on the order of unity ($\beta_{loc} \sim 1$). For a typical measurement, a KR angle of ~ 100 $\mu rad$ was observed at 10 K in $MoS_2$ strained by ~ 0.5% and in the presence of a current density of $J \sim 10$ A/m. Using Eqn. (1) and the experimental parameters we estimate the circular dichroism to be $A_+ - A_- \sim (2-3) \times 10^{-5}$ (KR angle $\theta \sim 20 - 30\ \mu rad$) if we take $\Gamma \sim 3.5$ meV (thermal broadening), $\Delta' \sim 2$ eV, $v_F \approx 0.6 \times 10^6$ ms$^{-1}$ (ref. [4]), $\alpha \sim 1$ (ref. [30, 31]) and peak absorbance $A \sim 0.3 - 0.5$ (Supplementary Fig. S11). We also estimate the valley magnetization to be $M_V \sim 4 \times 10^{-11}$ Amp. It corresponds to a volumetric converse magnetoelectric coefficient $\alpha_{zy} \sim \mu_0 M_V / \mathcal{E}_{dc} t \sim 0.5$ ps/m. Here $\mu_0$ and $t$ ($\approx 0.67$ nm) denote the vacuum permeability and the approximate thickness of monolayer $MoS_2$, respectively.

**Acknowledgments**

We acknowledge Wenyu Shan, Di Xiao, Inti Sodemann and Liang Fu for fruitful discussions. The research was supported by the National Science Foundation DMR-1420451 for sample and device fabrication and the US Department of Energy, Office of Basic Energy Sciences under award no. DESC0013883 and the Air Force Office of Scientific Research under grant FA9550-14-1-0268 for optical spectroscopy measurements. Support for data analysis and modeling was provided by the Air Force Office of Scientific Research under grant FA9550-16-1-0249 (K.F.M.) and the National Science Foundation DMR-1410407 (J.S.). This work was also supported by the National Research Foundation of Korea Grant funded by the Korean government (S2017A040300024) (J.L.) and a David and Lucille Packard Fellowship and a Sloan Fellowship (K.F.M.).

**Author contributions**

J.L., K.F.M. and J.S. conceived and designed the experiments, analyzed the data and co-wrote the manuscript. J.L., Z.W. and H.X. fabricated the devices and performed the measurements. All authors discussed the results and commented on the manuscript.

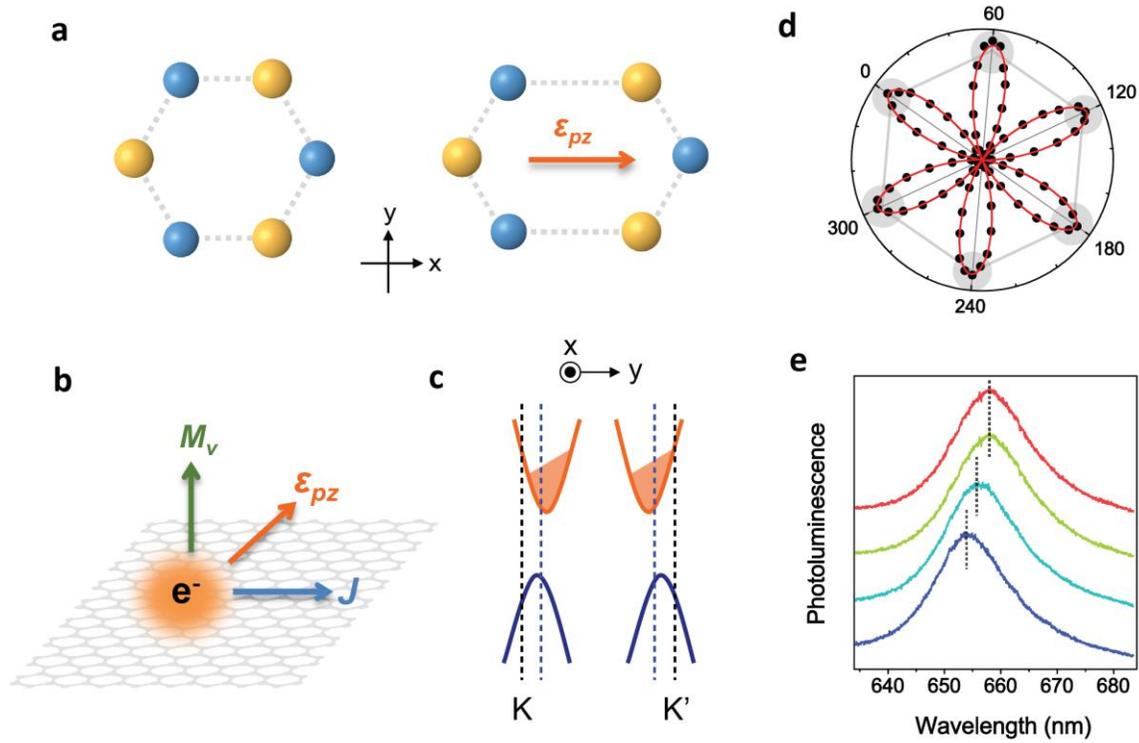

**Figure 1 | Valley magnetoelectric effect in strained single-layer MoS₂. a**, Piezoelectric field $\mathcal{E}_{pz}$ is produced (right) as a result of stress-induced distortion of the hexagonal lattice (left) and the associated ion charge polarization. Blue and yellow balls represent Mo and S atoms, respectively. **b**, Out-of-plane valley magnetization $M_V$ is induced by an in-plane current density $J$ in the presence of the piezoelectric field $\mathcal{E}_{pz}$. **c**, Electronic band structure of *n*-doped single-layer MoS₂ around the K and K' valleys of the Brillouin zone. The band extrema are shifted from the K/K' point (black dashed lines) with opposite direction by a uniaxial strain, which also do not coincide with the extrema of the Berry curvature distribution (blue dashed lines). The Fermi level is tilted under an in-plane bias electric field. **d**, Intensity of the second-harmonic component parallel to the excitation polarization ($I$) as a function of the excitation polarization angle measured from the armchair direction ($\theta$) (symbols). Solid line is a fit to $I = I_0\cos^2(3\theta)$ with $I_0$ denoting the maximum intensity. **e**, Photoluminescence (PL) spectrum of single-layer MoS₂ on a flexible substrate with 0% (blue), 0.1% (cyan) and 0.2% (green) strain along the armchair direction. Red line is the PL spectrum of the flake after being transferred onto a silicon substrate. The spectra are normalized to the peak intensity and vertically shifted.

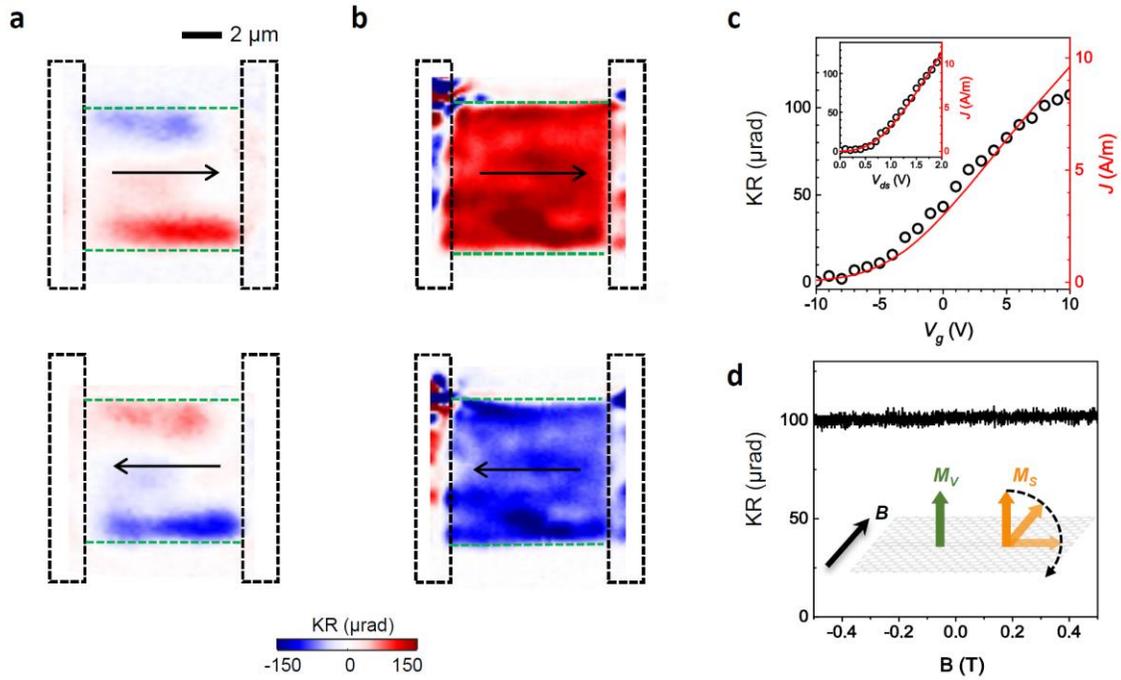

**Figure 2 | Valley Hall effect and valley magnetoelectric effect in single-layer MoS₂. a**, **b**, Kerr rotation image of an unstrained (**a**) and strained (**b**) single-layer MoS$_2$ device under two opposite bias directions (black arrows). Boundaries of the electrodes and the device channel are marked in dashed black and green lines, respectively. The unstrained device was measured with $V_{ds}$ = 2.5 V and $V_g$ = 0 V ($J$ = 22 A/m). KR is observed at channel edges only. The strained device was measured with $V_{ds}$ = 2.5 V and $V_g$ = 20 V ($J$ = 13 A/m). KR is observed over the entire channel. **c**, Gate dependence of the KR at a fixed location on the strained device (symbols) and gate dependence of channel current density $J$ measured with $V_{ds}$ = 2.5 V (red solid line). Inset shows the corresponding bias dependences with $V_g$ = 20 V. **d**, KR at a fixed location on the strained device as a function of in-plane magnetic field $B$. Inset illustrates the Hanle effect for spin magnetization $M_S$ and the absence of it for valley magnetization $M_V$.

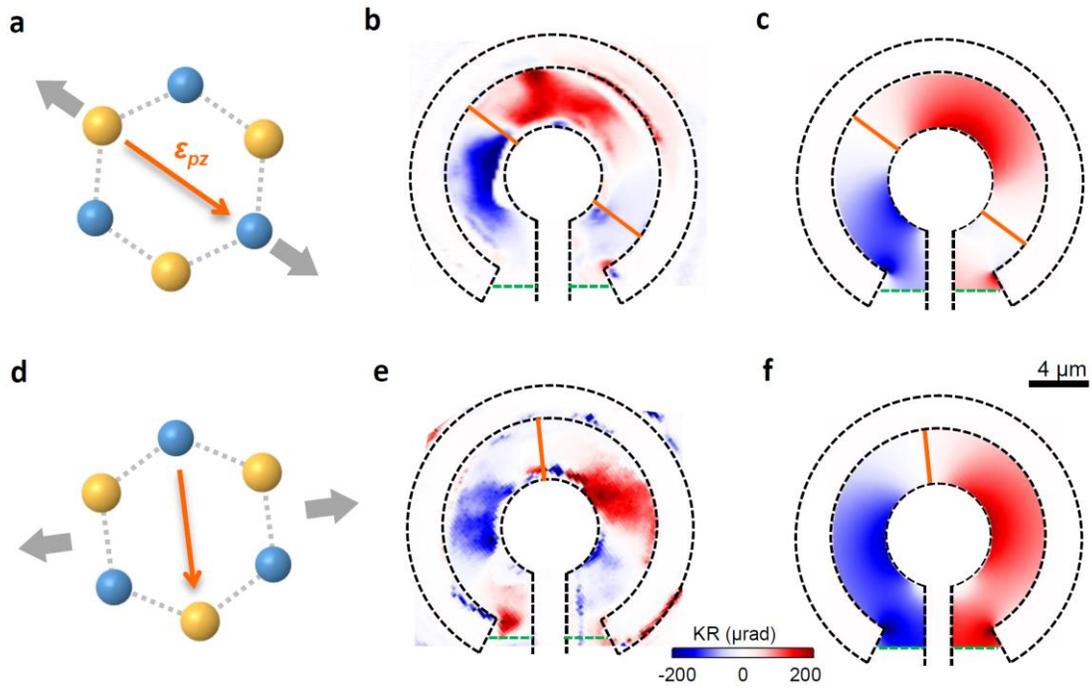

**Figure 3 | Dependence on current direction of the valley magnetoelectric effect. a**, Piezoelectric field $\mathcal{E}_{pz}$ is produced along the armchair direction when single-layer MoS$_2$ is stressed along the armchair direction; **b**, KR image of sample (**a**) measured in the Corbino disk geometry with $V_{ds}$ = 1 V and $V_g$ = 25 V ($J$ = 3.9 A/m); **c**, Predicted spatial distribution of valley magnetization by Eqn. (1). The direction of $\mathcal{E}_{pz}$ is shown in orange lines. Boundaries of the electrodes and the device channel are marked in dashed black and green lines, respectively. **d** to **f** are the same as **a** to **c** with single-layer MoS$_2$ stressed along the zigzag direction. The KR image was measured under $V_{ds}$ = 6.9 V and $V_g$ = 20 V ($J$ = 2.2 A/m).

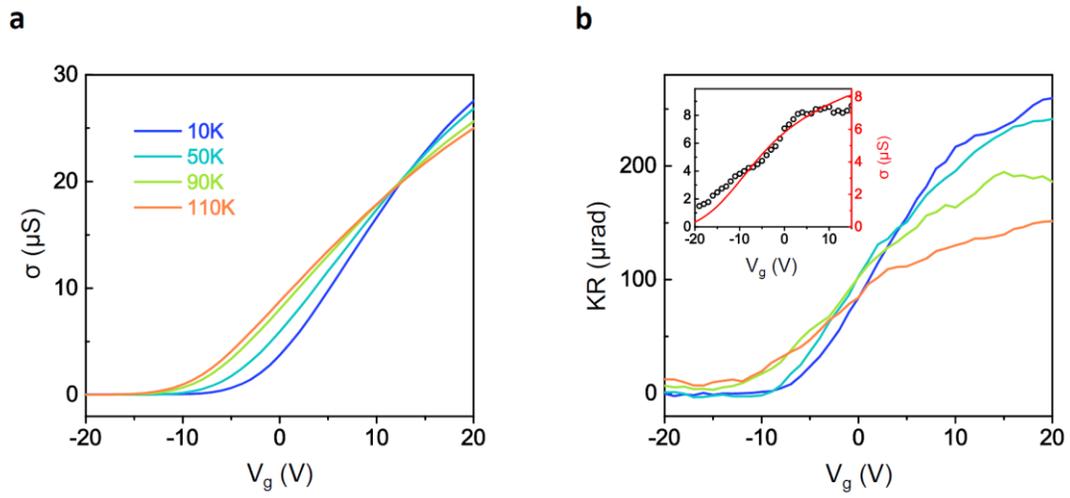

**Figure 4 | Temperature dependence of the valley magnetoelectric effect.** Gate dependence of the two-point conductance (**a**) and KR at a fixed location on a strained device (**b**) measured with $V_{ds}$ = 2 V at four selected temperatures 10, 50, 90, and 110 K. Inset of **b** displays the gate dependence of the KR (symbols) and conductance (red solid lines) at room temperature.